\documentclass[10pt, conference, compsoc]{IEEEtran}
\usepackage{times}
\usepackage{booktabs}
\usepackage{subfig}
\usepackage{float}
\usepackage{color}
\usepackage{graphicx}
\usepackage{amssymb}
\usepackage{amsthm}
\usepackage{mathtools}
\usepackage{enumitem}
\usepackage{listings}
\usepackage[numbers,sort]{natbib}
\usepackage{caption}
\usepackage{breqn}
\usepackage{framed}
\usepackage{algorithmic}
\usepackage{array}
\usepackage{mdwmath}
\usepackage{mdwtab}
\usepackage{url}
\usepackage{filecontents}

\setenumerate[1]{itemsep=0.5pt,partopsep=0pt,parsep=\parskip, topsep=2pt, leftmargin=12pt,}

\begin{document}

\title{Revisiting FPGA Acceleration of Molecular Dynamics Simulation \\ with Dynamic Data Flow Behavior in High-Level Synthesis}


\author{Jason Cong, Zhenman Fang, Hassan Kianinejad, Peng Wei 
\\Center for Domain-Specific Computing, University of California, Los Angeles
\\\{cong, zhenman, hasan, peng.wei.prc\}@cs.ucla.edu}

\maketitle
\pagestyle{plain}


\begin{abstract}

Molecular dynamics (MD) simulation is one of the past decade's most important tools for enabling biology scientists and researchers to explore human health and diseases. However, due to the computation complexity of the MD algorithm, it takes weeks or even months to simulate a comparatively simple biology entity on conventional multicore processors. The critical path in molecular dynamics simulations is the force calculation between particles inside the simulated environment, which has abundant parallelism. Among various acceleration platforms, FPGA is an attractive alternative because of its low power and high energy efficiency. However, due to its high programming cost using RTL, none of the mainstream MD software packages has yet adopted FPGA for acceleration. 

In this paper we revisit the FPGA acceleration of MD in high-level synthesis (HLS) so as to provide affordable programming cost. Our experience with the MD acceleration demonstrates that HLS optimizations such as loop pipelining, module duplication and memory partitioning are essential to improve the performance, achieving a speedup of 9.5X compared to a 12-core CPU. More importantly, we observe that even the fully optimized HLS design can still be 2X slower than the reference RTL architecture due to the common dynamic (conditional) data flow behavior that is not yet supported by current HLS tools. To support such behavior, we further customize an array of processing elements together with a data-driven streaming network through a common RTL template, and fully automate the design flow. Our final experimental results demonstrate a 19.4X performance speedup and 39X energy efficiency for the widely used ApoA1 MD benchmark on the Convey HC1ex FPGA compared to a 12-core Intel Xeon server.

\end{abstract}

\section{Introduction}
\label{sec:introduction}

Molecular dynamics (MD) simulation \cite{compiled_namd, ks_uiuc_website, NAMD_performance_STMV, grape, GRAPE_protein, anton_1, anton_2, anton_3, D_E_SHAW_website} is one of the past decade's most important tools in that it enables biology scientists and researchers to explore human health and diseases. In order to observe those critical biology phenomena, it iteratively simulates the motions of the molecular dynamics at an atomic level. At each iteration, it first calculates the forces applied for each atom and then updates the atom's motion. Each iteration, it simulates one or a few femtoseconds ($10^{-15}$ seconds) in real time, and usually it needs at least nanoseconds ($10^{6}$ iterations) or even milliseconds ($10^{12}$ iterations) for practical usage. This makes the MD simulation very computation-intensive, and conventional multicore processors fall far behind the computation requirements. 

Considering the importance and computation requirements of the MD simulation, there have been numerous efforts that try to accelerate it. We can divide the prior studies into the following three main categories: 1) ASIC-based acceleration; 2) GPU-based acceleration; and 3) FPGA-based acceleration. First, several ASIC-based machines have been built to accelerate the MD simulation, including MD-GRAPE \cite{grape}, GRAPE-based Protein Explorer \cite{GRAPE_protein}, and Anton from D.E. Shaw \cite{anton_1, anton_2, anton_3}. The ASIC-based approaches can achieve several orders-of-magnitude performance and energy gains. However, the MD algorithms keep evolving constantly due to the increased accuracy requirements. The nonrecurring engineering cost and design time of ASIC-based approaches pushes researchers and developers to look for alternative solutions. Therefore, the GPU-based acceleration (e.g., \cite{NAMD_GPU_1, NAMD_GPU_2, NAMD_GPU_3}) has gained much attention because GPUs are widely used in high-performance computing and provide much better programmability. Though GPUs can achieve high performance, they are very power-hungry.

Compared to ASIC-based and GPU-based acceleration approaches, FPGA-based acceleration emerges as an attractive alternative because FPGAs provide better programmability than ASICs, and better power and energy efficiency than GPUs. In prior studies, there have been several attempts (e.g., \cite{herbordt_1, herbordt_2, compiled_namd, dalian, khaled}) to accelerate MD using FPGAs. Most of these prior studies mainly focus on the acceleration for the critical path of the MD algorithm, which is the computation of short-range, non-bonded forces for each atom in every simulation iteration. Significant performance improvement and energy savings have been reported by these studies for the widely used ApoA1 MD benchmark \cite{NAMD}. However, they use an RTL design for the FPGA implementation, which leads to significant design time and engineering cost. As a result, none of the mainstream MD software packages have yet adopted FPGA for acceleration. 

In this paper we revisit the FPGA acceleration for MD simulation in high-level synthesis (HLS) \cite{journal_2011, vivado, altera_ocl, legup} that raises the abstraction level from RTL to C/C++. Our goal is achieve a comparable performance of prior reference RTL acceleration of MD simulation \cite{herbordt_1, herbordt_2}, but with more affordable programming cost in HLS. Similar to prior studies \cite{herbordt_1, herbordt_2, compiled_namd, dalian, khaled}, we focus on the acceleration of the critical path of the MD algorithm, that is, the computation of short-range, non-bonded forces for each atom, as explained in Section~\ref{sec:namd}. To provide more insights into the community, we start with a naive HLS design and optimize the HLS design all the way to generate the reference architecture written in RTL \cite{herbordt_1, herbordt_2}. In summary, this paper makes the following contributions.

\begin{enumerate}

\item We observe a limitation in current HLS tools --- HLS does not support the common dynamic (conditional) data flow behavior in high performance computing --- which can make some processing elements underutilized. This limitation makes the performance of the fully optimized HLS version of MD 2X below the reference RTL design. 

\item We provide a design automation flow to generate the enhanced architecture with dynamic data flow behavior from an HLS design, where we customize an array of processing elements together with a data-driven streaming network through a common RTL template.

\item Our final experimental results demonstrate a 19.4X performance speedup and 39X energy efficiency for the widely used ApoA1 MD benchmark on the Convey HC1ex FPGA compared to a 12-core Intel Xeon server. 

\end{enumerate}

The remainder of this paper is organized as follows. Section~\ref{sec:namd} describes essential background information on MD simulation. Section~\ref{sec:setup} describes the experimental setup used throughout this paper. Section~\ref{sec:init-hls} presents our experience with high-level synthesis (HLS) and breaks down the contribution of major HLS optimizations. Section~\ref{sec:observation} illustrates the observation of current HLS limitation---lack of support to generate the enhanced architecture with dynamic data flow behavior. Section~\ref{sec:beyond-hls} presents the design automation flow to generate the enhanced MD accelerator architecture from an HLS design. Section~\ref{sec:eval} presents the design space exploration and final results of the enhanced architecture. Finally, Section~\ref{sec:concl} concludes this paper.

\section{MD Algorithm and Its Code Transformation}
\label{sec:namd}

In this section we briefly introduce the MD algorithm and necessary code transformations to make it applicable for HLS-based FPGA acceleration.

\subsection{MD Algorithm}

\begin{figure}[!t]
\centering
\includegraphics[scale=0.4]{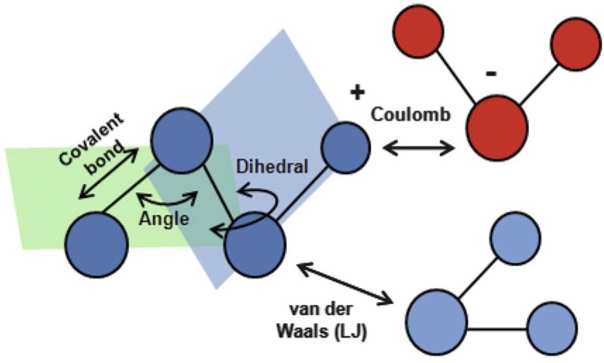}
\caption{Major forces in MD simulation~\cite{herbordt_1}, including bonded and non-bonded forces (van der Waals Leonard-Jones and Coulomb forces).}
\vspace{-0.1in}
\label{fig:forces}
\end{figure}

Molecular dynamics (MD) simulation \cite{compiled_namd, ks_uiuc_website, NAMD_performance_STMV, grape, GRAPE_protein, anton_1, anton_2, anton_3, D_E_SHAW_website} is one of the most important tools for observing those critical biology phenomena. Basically, it simulates the motions of the molecular systems at an atomic level by $10^{6}$ to $10^{12}$ iterations for practical usage, which makes it very time consuming. At each iteration, it first calculates the forces applied for each atom and then updates the atom's motion. Figure~\ref{fig:forces}~\cite{herbordt_1} describes the major forces applied on each atom, including bonded and non-bonded forces. Among all the force computations, the most time consuming part is the non-bonded force computation that includes the van der Waals (Leonard-Jones) force calculation and the Coulomb force calculation between every atom pair. Equations 1 and 2 describe the formula for calculating the Leonard-Jones (LJ) and Coulomb (CL) force, respectively. In both equations, r represents the distance vector between two atoms, and the parameters k, q1, q2, A, B are all constants.  

\begin{equation}
	F(LJ) = (A/|r|^{14} - B/|r|^8) * r
\end{equation}

\begin{equation}
	F(CL) = k * q1 * q2 / |r|^3 * r
\end{equation}

\begin{figure}[!ht]
\centering
\includegraphics[scale=0.35]{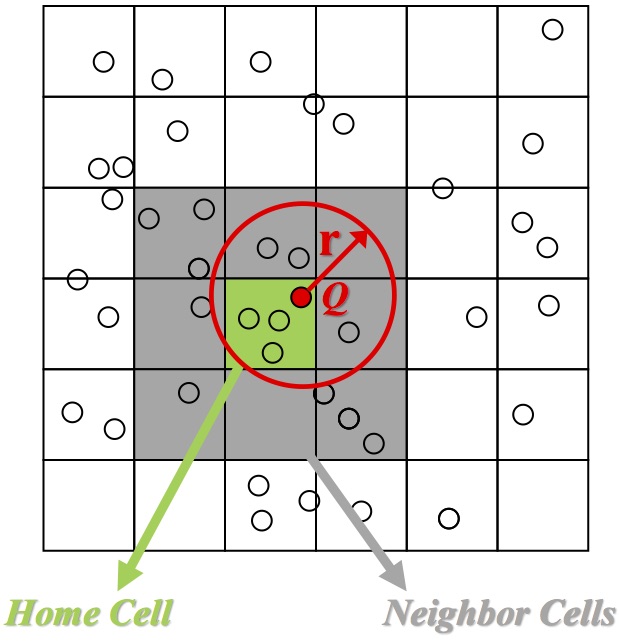}
\caption{Simulation space for atom \emph{Q} in \textit{home cell} where its two-dimensional \textit{neighbor cells} are shown in grey. Cells have edge size roughly equal to the cutoff radius. The cut-off circle for atom \emph{Q} is also shown.}
\vspace{-0.15in}
\label{fig:cell-list}
\end{figure}

As observed in Equations 1 and 2, both forces will become smaller and have less impact on the motion of the atom when the distance vector r increases. Therefore, a common way to reduce the time-consuming, non-bonded force computation is to apply a cutoff. It only calculates the non-bonded forces using Equations 1 and 2 if the distance vector between two atoms is within a cutoff radius (i.e., short-range, non-bonded forces). Otherwise, if it is beyond the cutoff radius, it uses some other approximations to further reduce the computation complexity from O($N^2$) to O(N) for the Coulomb force and ignore the Leonard-Jones force as 0. 

As an initial effort, we mainly focus on the acceleration of the short-range, non-bonded force computation that is the critical path of the MD algorithm, similar to prior studies \cite{herbordt_1, herbordt_2, compiled_namd, dalian, khaled}. Therefore, we will only report the performance of one simulation iteration for the short-range, non-bonded force computation throughout this paper. We do not consider the long-range Coulomb force for FPGA acceleration and leave it to CPU, since its computation complexity can be reduced to O(N) and it is not in the critical path. Without loss of generality, we use floating-point for both our CPU baseline and FPGA accelerators for a fair comparison.

Calculating the short-range, non-bonded force for each atom involves accumulating the force contribution of all atoms within the cutoff radius. One well-known method that takes advantage of the cutoff radius to further reduce the computation is the cell-list-based approach \cite{herbordt_1, herbordt_2}, which is described in Figure~\ref{fig:cell-list}. It divides the whole simulation space into cell cubes where the edge-length of each cell roughly equals the cutoff radius. As a result, the short-range, non-bonded forces of an atom \emph{Q} in its \textit{home cell} can only be applied by atoms in its 3*3*3 neighbor cells (including the home cell itself). Using the cell-list-based approach, we summarize the core computation of the MD critical path, as shown in Algorithm 1.

\vspace{0.1in}

\noindent \textbf{Algorithm 1}. Core computation code in the MD critical path (a single simulation iteration).
\begin{framed}
\setlength{\parindent}{0pt}
\textbf{foreach} \textit{cell} \textbf{do} \\
\setlength{\parindent}{10pt}
\indent \textbf{foreach} \textit{neighbor cell} \textbf{do} \\
\indent \indent \textbf{foreach} \textit{atom} in \textit{home cell} \textbf{do} \\
\indent \indent \indent total force = 0; \\
\indent \indent \indent \textbf{foreach} \textit{atom} in \textit{neighbor cell} \textbf{do} \\
\indent \indent \indent \indent \textbf{if} (distance() \textless  cutoff radius) \\
\indent \indent \indent \indent \indent compute force(); \\
\indent \indent \indent \textbf{end foreach} \\
\indent \indent \textbf{end foreach} \\
\indent \textbf{end foreach} \\
\textbf{end foreach} 
\end{framed}

\vspace{0.1in}

\subsection{Code Transformation for HLS}

In this paper we focus on the acceleration of the critical path of MD simulation, as shown in Algorithm 1. Though a number of well-known MD software packages, such as NAMD~\cite{NAMD} and NAMD-Lite~\cite{NAMD_lite}, have been developed and widely used by the community, we find that it is difficult to accelerate them on FPGA using HLS directly. Two major code transformations are needed to make these software C/C++ codes compatible with HLS. Since they are quite common for high-performance computing (HPC) application, we would like to share our code transformation experience with the community. 

\begin{enumerate}

\item The MD algorithm heavily uses C structures to represent the position and properties (such as charge and mass) for each atom in the software source code. However, current FPGA HLS tools have limited support for C structures that are passed through functions as function parameters: HLS tools will decompose those C structures into individual data members that may not be what users want for interface ports. Therefore, we have to rewrite these structure arrays into the C primitive data arrays that are correctly mapped onto the interface ports manually. We will investigate compiler techniques to automatically handle the source-code level transformation in our future work. 

\item The MD algorithm also uses some pointer-based data structures to efficiently organize the atoms. For example, as the number of atoms inside each cubic cell (shown in Figure~\ref{fig:cell-list}) is not fixed, they are structured in a linked-list in the software source code. However, current FPGA HLS tools do not support complex pointer operations (linked-list, graph, tree, etc.) that are common in the software. Hence, we have to rewrite these complex pointer-based data structures into simple arrays. This code transformation is very difficult to achieve automatically, and we manually make all the necessary transformations.

\end{enumerate}

We use the source code for Algorithm 1 with the above transformations as our baseline. And we will release our baseline and HLS-based accelerator code used throughout this paper, once the paper is accepted for publication.

\section{Experimental Setup}
\label{sec:setup}

Before starting the acceleration of the critical path of the MD algorithm, we describe the experimental setup used throughout this paper. The baseline CPU platform we use for comparison is a two-socket 12-core Intel Xeon E5-2640 (32nm) server running at 2.5GHz. The total power consumption of the server is 300W. For a fair comparison, we implement the multithread version of our simplified MD algorithm using OpenMP.

\begin{table}[!ht]
\renewcommand{\arraystretch}{1.3}
\caption{FPGA platform information.}
\label{FPGA_platform_information}
\centering
\begin{tabular}{|c|c|}
\hline
Platform & Convey HC-1ex\\
\hline
FPGA fabric & Virtex-6 VLX760 x 4
\\
\hline
Logic Cell&474K x 4
\\ 
\hline
DSP&864 x 4\\ 
\hline
BRAM&26Mb x 4\\ 
\hline
DDR BW&80GB/s\\ 
\hline
DDR size&64GB\\ 
\hline
Power&150W\\ 
\hline
Technology&40nm\\ 
\hline
\end{tabular}
\end{table}

\begin{figure*}
\centering
\includegraphics[scale=0.2]{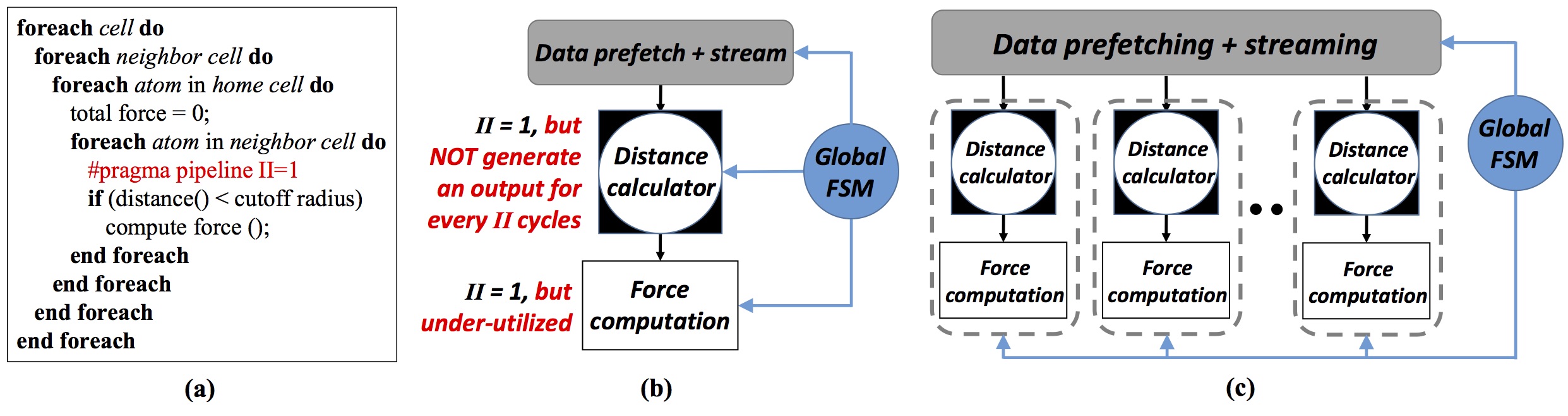}
\caption{MD accelerator generated by HLS: (a) HLS code of Algorithm 1 that pipelines the innermost loop with II=1; (b) single processing element generated by HLS for (a); (c) Architecture generated by HLS for (a) with module duplication.}
\label{fig:hls_arch}
\end{figure*}

\begin{table*}
\renewcommand{\arraystretch}{1.3}
\caption{HLS results on CPU and FPGA.}
\label{Initial_HLS_results}
\centering
\begin{tabular}{|c|c|c|c|c|c|}
\hline
Platform & Parallelism & Run time & Speedup & Power &Energy Efficiency\\
\hline
Intel Server & 1 core & 33 sec & - & - & -\\
\hline
Intel Server & 12 cores & 3.5 sec & 1 & 300W & 1\\
\hline
Convey HC1ex & 4 \textit{force pipelines} x 4 & 0.37 sec & 9.5X & 150W & 19X \\
\hline
\end{tabular}
\vspace{-0.1in}
\end{table*}

We use the Convey HC1ex machine~\cite{convey_website} for the acceleration, which is a CPU and FPGA coupled platform for high-performance computing. It consists of two Intel Xeon CPUs and four Xilinx Virtex-6 VLX760 FGPA chips running at 100 MHz. It also provides a high-performance interface to the DRAM system, with up to 80GB/s bandwidth. In our experiments we only use FPGAs in the Convey machine, and the power consumption is 150W. The detailed information of the Convey HC1ex machine is summarized in Table~\ref{FPGA_platform_information}. We use the Xilinx Vivado HLS tool \cite{vivado} to synthesize our HLS designs. 

There are many benchmarks available for the MD software. We choose \textit{ApoA1} to evaluate our accelerator design because it has been the standard MD cross-platform benchmark for years \cite{NAMD}. It contains about 100,000 atoms scattered among 700 cubic cells in three dimensions.

\section{HLS Acceleration of MD}
\label{sec:init-hls}

High-level synthesis (HLS) has been a great success in reducing design time and engineering cost for accelerator designs because HLS raises the abstraction level from RTL to C/C++ \cite{journal_2011, vivado, altera_ocl, legup}. Nevertheless, mapping the application C/C++ code to the desired architecture in HLS is still challenging. HLS will yield poor performance if the application developer does not have enough knowledge of the FPGA architecture or not employ optimization techniques carefully. 

In this paper we accelerate the MD critical path using HLS so as to provide affordable programming cost. We will first explain the architecture generated by HLS and then break down the performance improvement and energy savings using different HLS optimizations including loop pipelining, module duplication, and memory optimizations. 

\subsection{HLS Design}

To generate an optimized MD accelerator for Algorithm 1, major well-known HLS optimizations are applied to pipeline the innermost loop that calculates the distance and computes the force. The goal is to achieve pipeline initialization interval (II) of 1 in all FPGA designs, i.e., in each cycle, the pipeline can process one input data. The applied HLS optimizations include loop unrolling and pipelining, and memory optimizations like data reuse optimization. Figure~\ref{fig:hls_arch}(b) presents the single processing element generated by HLS for the code shown in Figure~\ref{fig:hls_arch}(a). It mainly has three modules: 1) data prefetching and streaming module, 2) distance calculator module, and 3) force computation module. Duplicating the processing elements inside the innermost loop generates the architecture shown in Figure~\ref{fig:hls_arch}(c). 

\subsection{HLS Results}
\label{sec:hls-results}

We present the evaluation results for the baseline CPU and fully optimized HLS on the Convey FPGA as shown in Table \ref{Initial_HLS_results}. First, the 12-core CPU version can achieve a 9.4X performance speedup compared to the single-core CPU. It cannot achieve close to 12X speedup due to the imbalanced computation for each core caused by the dynamic data flow behavior explained in Section~\ref{sec:observation}. To be fair, we will use this multicore result as a baseline for comparison. Second, on the Convey FPGA platform, the HLS-generated MD accelerator achieves 9.5X performance speedup and 19X energy efficiency compared to the 12-core baseline.

To get a better idea of the performance and energy contribution of each major HLS optimization, we further illustrate the performance speedup and energy efficiency breakdown of HLS optimizations on the Convey FPGA platform, compared to the Intel 12-core Xeon server. As shown in Table~\ref{different_FPGA_acc}, a naive HLS design actually degrades the performance by 100X and increases the energy consumption by 50X compared to the 12-core server. One major HLS optimization that can significantly improve the performance of HLS is loop pipelining (together with loop unrolling within the innermost loop). Loop pipelining bridges the gap of HLS and achieves similar performance of the 12-core server, while saving 2X energy. The second major HLS optimization is module duplication together with necessary memory partitioning optimizations. Table~\ref{FPGA_resource_utilization_for_MD_accelerator} summarizes the FPGA resource consumption for a single accelerator module on the Convey FPGA. As a result, we can duplicate 4 modules per FPGA chip, resulting in 16 modules for the 4 FPGA chips. As shown in Table~\ref{different_FPGA_acc}, this fully optimized HLS can achieve a 9.5X performance speedup and 19X energy efficiency compared to the 12-core server. 

\begin{table}
\renewcommand{\arraystretch}{1.3}
\caption{Performance speedup and energy efficiency breakdown of major HLS optimizations on Convey FPGA platform compared to the Intel 12-core Xeon server.}
\label{different_FPGA_acc}
\centering
\begin{tabular}{|l|l|l|l|}\hline
Platform&Run time&Speedup&Energy Efficiency\\ \hline
Intel 12-core server&3.5 sec&1&1 \\ \hline
Naive HLS result&338 sec&0.01X&0.02X \\ \hline
HLS result with &3.52 sec&1.0X&2X \\
pipeline (single module) &&& \\ \hline 
Fully optimized HLS &0.37 sec&9.5X&19X \\
(module duplication) &&& \\ \hline 
\end{tabular}
\end{table}

\begin{table}
\renewcommand{\arraystretch}{1.3}
\caption{FPGA resource utilization for one MD accelerator in Convey FPGA platform.}
\label{FPGA_resource_utilization_for_MD_accelerator}
\centering
\begin{tabular}{|c|c|c|c|}
\hline
BRAM 18K & DSP48E & FF & LUT \\
\hline
153 (10\%) & 222 (23\%) & 26617 (2\%) & 31144 (6\%) \\
\hline
\end{tabular}
\end{table}

\section{Analysis of HLS Limitation}
\label{sec:observation}

Although HLS-based acceleration achieves significant performance speedup and energy efficiency as explained in Section~\ref{sec:hls-results}, we still observe one source of inefficiency in the architecture generated by HLS caused by the dynamic (conditional) data flow behavior that is common in high-performance computing applications. In this paper we will use the MD accelerator as an example to illustrate the inefficiency of the HLS-generated accelerator, and explain the limitation of existing HLS tools to generate the enhanced reference accelerator.

\subsection{Inefficiency of HLS-Generated Accelerator}

While HLS tools are very successful in synthesizing static control flows in regular applications, they usually yield inefficient results for applications with a dynamic data flow behavior where conditional execution exists within a processing element. 

In the HLS-generated accelerator architectures, there are two processing elements (PEs), i.e., the distance calculator PE and force computation PE, which are connected to each other through a FIFO interface, as shown in Figure~\ref{fig:hls_arch}(b) and Figure~\ref{fig:hls_arch}(c). Each PE is pipelined with II (initiation interval) as 1, such that every clock cycle it reads the inputs, computes the result and writes the output. However, as described earlier in Section~\ref{sec:namd}, atom pairs farther from the cutoff radius have much less impact on each other in the force computation. To be more specific, for each atom in the \textit{home cell}, only \textit{partner atoms} inside the sphere around it (shown in Figure~\ref{fig:cell-list}) pass the \textbf{if} condition wrapping around distance calculator as shown in Figure~\ref{fig:hls_arch}(a). Caused by this conditional execution, the producer PE (i.e., distance calculator PE) does not generate an output each clock cycle. As a result, the consumer PE (i.e., the force computation PE) that expects an input and computation every clock cycle will be underutilized.

\subsection{The Enhanced Accelerator Design and The Limitation of HLS to Generate It}

\begin{figure}[!t]
\centering
\includegraphics[scale=0.25]{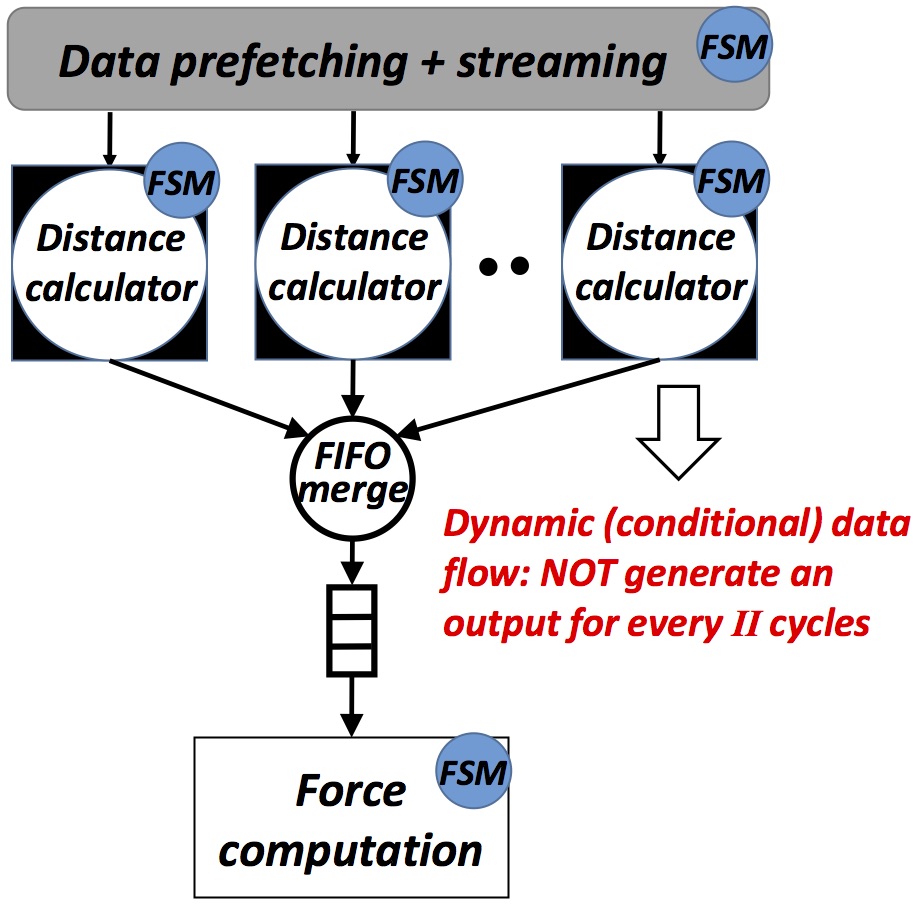}
\caption{Enhanced MD accelerator architecture for more efficient module utilization with dynamic data flow in streaming network.}
\vspace{-0.1in}
\label{fig:revised_arch}
\end{figure}

\begin{table*}[!t]
\renewcommand{\arraystretch}{1.3}
\caption{FPGA resource utilization for a single copy of the enhanced architecture in Convey platform.}
\label{Area_breakdown}
\centering
\begin{tabular}{|l|c|c|c|c|}
\hline
Module & BRAM 18K & DSP48E & FF & LUT \\\hline
Data prefetching + streaming & 265 (18\%) & 30 (3\%) & 5913 ($\sim 0$\%) & 5671 (1\%)  \\\hline
Distance calculator & 0 (0\%) & 19 (2\%) & 2295 ($\sim 0$\%) & 2052 ($\sim 0$\%) \\\hline
Force computation & 4 ($\sim 0$\%) & 173 (20\%) & 16845 (1\%) & 16920 (3\%) \\\hline
\end{tabular}
\end{table*}

To alleviate the under-utilization problem analyzed above, a multi-producer one-consumer data flow is expected. Considering the fact that the distance calculation module consumes much less FPGA computing resources than the force computation module, as shown in Table~\ref{Area_breakdown}, we can take advantage of this opportunity to take a more efficient approach by duplicating the distance calculation module. The enhanced multi-producer-single-consumer accelerator architecture is shown in Figure~\ref{fig:revised_arch}, which is similar to the reference architecture written in RTL \cite{herbordt_1, herbordt_2}.

However, most of the existing HLS tools (including Xilinx Vivado HLS) assume a Kahn Process Network (KPN) model \cite{kpn_1, kpn_2} to synthesize C/C++ code to RTL design. That is, it assumes that every input to an PE produces an output in a data flow, i.e., there is no conditional execution in the data flow. 
The dynamic (conditional) data flow behavior of the architecture described in Figure~\ref{fig:revised_arch} violates this assumption and goes beyond the KPN model due to the conditional execution of the producer PE (i.e., distance calculator PE).
Therefore, we cannot easily generate the desired enhanced accelerator design using existing HLS tools like Xilinx Vivado HLS.

\section{Code Transformation and Automation to Generate the Enhanced Accelerator}
\label{sec:beyond-hls}

To automatically generate the enhanced accelerator design with dynamic data flow behavior, we present a joint HLS-RTL design in this section. At a high level, all the processing elements (PEs) are still generated by HLS and only the dynamic data flow is described in an RTL template. Next we present necessary code transformations and automation support to achieve this goal.

\subsection{Joint HLS-RTL Design and Code Transformations for Dynamic Data Flow Behavior} \label{sec:hls-rtl}

First of all, we use an RTL template shown in Algorithm 2 as a top module to describe the dynamic data flow behavior with conditional execution between the producer and consumer PEs. In our enhanced MD accelerator design example, PE0 represents distance calculator PE and PE1 represents force computation PE, and both PEs are still designed in HLS. In the RTL template, PE0 and PE1 are connected through FIFOs, and users can specify the desired copies of each PE as well. 

Moreover, mapping the application kernel (i.e., the distance calculator and force computation PEs) into this enhanced architecture is not straightforward using HLS. The reason is that each module now has its own independent finite state machine (FSM) to control the processing logic instead of a global FSM in the previous HLS-generated architecture shown in Figure~\ref{fig:hls_arch}. Therefore, the HLS coding style for each module needs to be changed. 

\vspace{0.15in}

\noindent \textbf{Algorithm 2}. Top-level RTL module template to describe the dynamic data flow behavior between two processing elements (PEs). In the MD accelerator, PE0 represents distance calculator PE and PE1 represents force computation PE.
\begin{lstlisting}
module md_accelerator 
    #(parameter NUM_PE0, NUM_PE1)
(/* input/output ports */)

// K: the number of PE0s that feeds a PE1
localparam K = NUM_PE0 / NUM_PE1;

generate
  genvar i;
  // generate PE0s
  for (i=0; i<NUM_PE0; i++)
  begin: PE0
    pe_0(
      // ports to FIFO
      // ...);
  end
  // generate PE1s
  for (i=0; i<NUM_PE1; i++)
  begin: PE1
    pe_1(
      // ports from FIFO
      // ...);
  end
  // generate FIFOs
  for (i=0; i<NUM_PE1; i++)
  begin: FIFO
    fifo(
      // ports from i*K to (i+1)*K PE0s
      // ports to i-th PE1
    )
  end
endgenerate
endmodule
\end{lstlisting}

\vspace{0.1in}

Algorithm 3 shows the HLS code changes that are needed in order to generate the enhanced architecture. All processing elements (PEs) are connected to each other through FIFOs in this approach. And these PEs will be instantiated with arbitrary duplication factors in the top RTL module described in Algorithm 2. It is worthwhile to mention that all PEs are running in parallel; and since they have their own local FSM, the pipeline setup time is cut down compared to the architecture generated by HLS in Figure~\ref{fig:hls_arch} where the global FSM enforces sequential execution between modules.

\vspace{0.2in}
\noindent \textbf{Algorithm 3}. Code transformations of each PE needed to generate the enhanced accelerator architecture.
\begin{framed}
\setlength{\parindent}{0pt}
\textbf{void} \textit{distance\_calculator} () \{ \\
\setlength{\parindent}{10pt}
\indent \textbf{while} (true) \{ \\
\indent \indent \emph{\#pragma pipeline II=1} \\
\indent \indent	read input FIFO(); \\
\indent \indent calculate\_distance(); \\
\indent \indent \textbf{if} (distance \textless cutoff radius) \\
\indent \indent \indent write output FIFO(); \\
\indent \} \\
\} 
\\
\textbf{void} \textit{force\_computation} () \{ \\
\setlength{\parindent}{10pt}
\indent \textbf{while} (true) \{ \\
\indent \indent \emph{\#pragma pipeline II=1} \\
\indent \indent	read input FIFO(); \\
\indent \indent compute\_force(); \\
\indent \indent write output FIFO(); \\
\indent \} \\
\}
\end{framed}

\vspace{0.1in}

\subsection{Design Automation Flow}

\begin{figure}[!ht]
\centering
\includegraphics[scale=0.32]{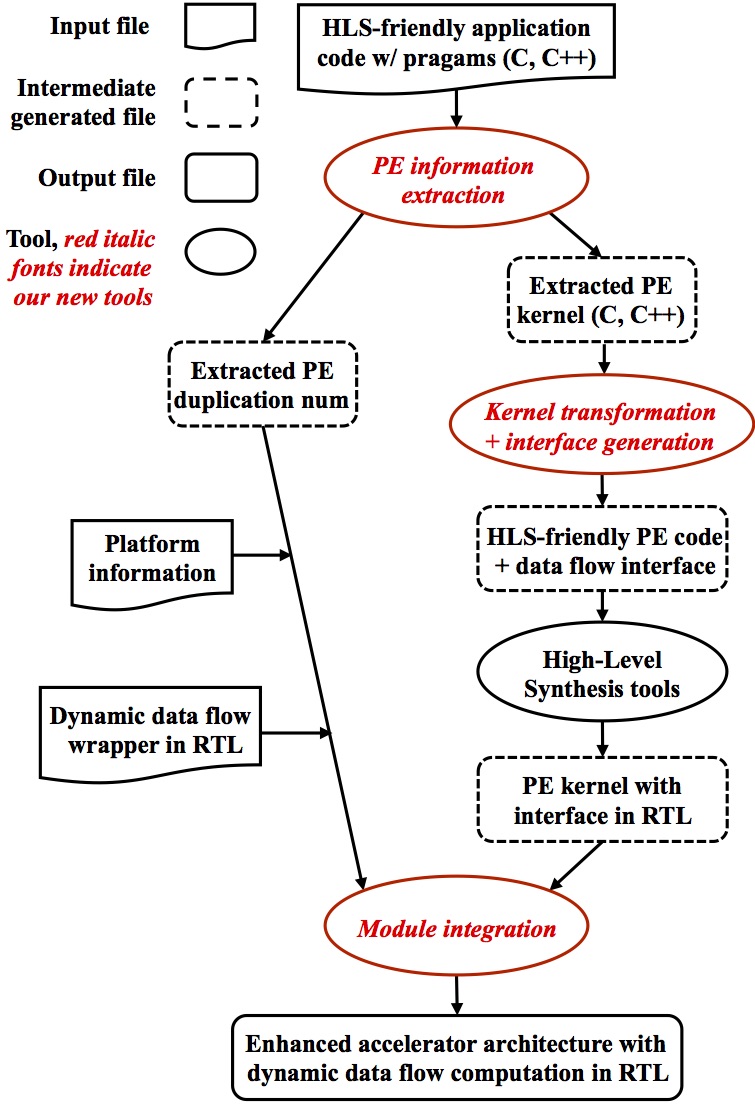}
\caption{Design automation flow to generate the enhanced accelerator architecture with dynamic data flow behavior.}
\vspace{-0.1in}
\label{fig:automation}
\end{figure}

To further reduce the programming cost, we also develop a design automation flow to generate the complete processor array architecture enhanced for dynamic data flow behavior. Figure~\ref{fig:automation} presents an overview of the automation flow. Programmers only need to write the code in HLS-friendly C or C++ language with a minimum number of \textit{pragmas} to indicate which function should be mapped to a processing element (\textit{PE}) and the duplication number. On one hand, our flow will automatically extract the pure computation core of the functions, and transform the extracted software kernel into the HLS-friendly C/C++ PE code that satisfies our data-flow-driven architecture. At the same time, our flow will also generate the associated data stream interface (i.e., FIFOs). After these HLS-friendly C/C++ files are generated, we call on the HLS tool (i.e., Xilinx Vivado HLS in this paper) to produce the RTL computation kernels for each PE with the data stream interface. On the other hand, by combining platform (FPGA) information with duplication ratios indicated by the user, we finally integrate all PE modules using our RTL wrapper template presented in Section~\ref{sec:hls-rtl}, to generate the enhanced processor array architecture with dynamic data flow behavior.

\section{Results of Enhanced Accelerator}
\label{sec:eval}

In this section we first decide the right duplication factor of the distance calculator PE in the enhanced accelerator through design space exploration. Then we present the final performance speedup and energy savings.

\subsection{Design Space Exploration}
\label{sec:dse}

To decide the duplication ratio of the \textit{distance calculator} and \textit{force computation} modules in the enhanced architecture, we do more design space exploration considering both the achieved speedup and area overhead (i.e., resource consumption overhead). One can also perform profiling on the software to see how frequently the \textbf{if} condition inside a \textit{PE} generates an output to the next \textit{PE}. First, we present the FPGA resource utilization (i.e., area) breakdown for a single copy of each module described in Figure~\ref{fig:revised_arch}. As shown in Table \ref{Area_breakdown}, DSP block usage is the dominating factor in all the resource consumption because the MD accelerator uses floating point computation. As a result, we mainly consider the DSP block usage when calculating the area overhead. We also find that the distance calculation module requires far less resources/area than the force computation module, as shown in Table \ref{Area_breakdown}. As a result, it is acceptable to duplicate more distance calculator modules, unlike the work in~\cite{cgpa}.

\begin{table}[!h]
\renewcommand{\arraystretch}{1.3}
\caption{Design space exploration: performance speedup and area overhead with different number of duplicating distance calculation modules and one single computation module on Convey FPGA platform.}
\label{pe_0_dup}
\centering
\begin{tabular}{|c|c|c|c|c|}\hline
Duplication&Execution time&Speedup&Area overhead&Speedup/area \\ \hline
1&3.5 seconds&1X&0&- \\ \hline
2&2.1 seconds&1.7X&8.7\%&1.56 \\ \hline
3&1.64 seconds&2.13X&17.4\%&1.81 \\ \hline
4&1.4 seconds&2.5X&25\%&2.0 \\ \hline
5&1.4 seconds&2.5X&33\%&1.88 \\ \hline
\end{tabular}
\end{table}

\begin{table*}
\renewcommand{\arraystretch}{1.3}
\caption{Final performance speedup and energy savings on Convey FPGA compared to 12-core Intel server.}
\label{final_results}
\centering
\begin{tabular}{|c|c|c|c|c|c|}
\hline
Platform and design & Parallelism & PE Utilization & Run time & Speedup & Energy Efficiency\\
\hline
Intel Xeon Server (Baseline) & 12 cores & - & 3.5 sec & 1 & 1\\
\hline
Convey HC1ex HLS Design  & 4 \textit{force pipelines} x 4 & around 25\% & 0.37 sec & 9.5X & 19X \\
\hline
Convey HC1ex HLS-RTL Design  & 2 \textit{force pipelines} x 4 & close to 100\% & 0.18 sec & 19.4X & 39X \\
\hline
\end{tabular}
\end{table*}

Table~\ref{pe_0_dup} presents the performance speedup, area overhead, and performance per area for duplicating different ratios of the distance calculation module and force computation module. The baseline we use here has only one single distance calculation and force computation module. Then we fix the number of force computation modules to one and increase the number of distance calculation modules from one to five. As shown in Table~\ref{pe_0_dup}, the performance increases until we have four duplicates of the distance calculation module. Interestingly, it also has the best performance per area. As a result, we choose four distance calculation modules connected to one force computation module in our final MD accelerator design. It has a 2.5X performance speedup and 25\% area overhead compared to the single copy of a distance calculation and force computation module.

\subsection{Final Results}

To further improve the performance of the enhanced MD accelerator, we can duplicate the enhanced design shown in Figure~\ref{fig:revised_arch}. For each FPGA chip on the Convey FPGA, the enhanced design will be duplicated by 2X due to the increased area overhead by the duplicated distance calculation modules (explained in Section~\ref{sec:dse}). Note that we cannot make 3X duplication because usually the timing cannot be met during the RTL synthesis when more than 90\% of the FPGA resources are used. It is also interesting to note that in our fully optimized HLS version we have four force computation modules per FPGA chip. However, as shown in the Table~\ref{final_results}, our enhanced accelerator with joint HLS-RTL design that provides the support of dynamic data flow behavior can still achieve a 2X better performance compared to the fully optimized HLS solution. The reason is that the force computation PE is underutilized (only around 25\% utilization) in the fully optimized HLS design, which is caused by the dynamic data flow behavior. While in the enhanced accelerator design, we address the dynamic data flow issue and the force computation PE is almost fully utilized. Compared to the 12-core CPU server, our final enhanced accelerator achieves a 19.4X performance speedup and 39X energy efficiency.

\section{Conclusion}
\label{sec:concl}

In this paper we revisited the acceleration of the critical path of the MD algorithm on FPGA using high-level synthesis (HLS). Our experience demonstrated that a naive HLS design actually degrades the performance, and HLS optimizations such as loop pipelining, module duplication and memory partitioning are essential in order to achieve high performance. Furthermore, we observed that the common dynamic (conditional) data flow behavior is not supported by current HLS tools and would limit the accelerator performance by around 2X. To address this issue, we provided an RTL template and an automated flow to customize an array of processing elements together with a data-driven streaming network, while users can still design their accelerators in HLS. We believe this design automation flow can be applied to other applications with dynamic data flow behavior as well. Our final experimental results demonstrated a 19.4X performance speedup and 39X energy efficiency for the widely-used ApoA1 MD benchmark on the Convey HC1ex FPGA compared to a 12-core Intel Xeon server.


\bibliographystyle{IEEEtran}
\small
\bibliography{arXiv16}

\end{document}